\documentclass{aa}
\usepackage{graphicx} %
\usepackage{upgreek} %
\usepackage{txfonts}

\usepackage{natbib}
\usepackage{hyperref}
\hypersetup{
    colorlinks=true,
    linkcolor=blue,
    filecolor=blue,      
    urlcolor=blue,
    citecolor=blue
}
\bibpunct{(}{)}{;}{a}{}{,} 

\newcommand{\sect}[1]{\text{Sect.~\ref{#1}}}
\newcommand{\fig}[1]{\text{Fig.~\ref{#1}}}
\newcommand{\tab}[1]{\text{Table~\ref{#1}}}

\newcommand{\kms}{\mathrm{km\,s^{-1}}}
\newcommand{\teff}{T_{\mathrm{eff}}}
\newcommand{\logg}{\log{g}}
\newcommand{\feh}{\mathrm{[Fe/H]}}
\newcommand{\dex}{\mathrm{dex}}
\newcommand{\imbalance}{$\Delta_{\textsc{I}-\textsc{II}}\,$}
\newcommand{\tife}{\mathrm{[Ti/Fe]}}
\newcommand{\tih}{\mathrm{[Ti/H]}}
\newcommand{\vturb}{\xi_{\mathrm{t}}}
\newcommand{\abund}{\mathrm{A\left(Ti\right)}}

\begin{document}

\title{ Titanium abundances in late-type stars}
\subtitle{II. Grid of departure coefficients and 
application to a sample of $70\,000$ stars}
\author{J.~W.~E.~Mallinson\inst{\ref{su}}
\and
K.~Lind\inst{\ref{su}}
\and
A.~M.~Amarsi\inst{\ref{uu}}
\and
K.~Youakim\inst{\ref{su}}}
    \institute{\label{su}Department of Astronomy, Stockholm
University, AlbaNova University Centre, Roslagstullsbacken, SE-106 91 Stockholm, Sweden\\
\email{jack.mallinson@astro.su.se}
\and
\label{uu}Theoretical Astrophysics, Department of
Physics and Astronomy, Uppsala University, Box 516,
751 20, Uppsala, Sweden}

   \date{}

 
  \abstract
   {Rapidly growing datasets from stellar spectroscopic surveys are providing unprecedented opportunities to analyse the chemical evolution history of our Galaxy. However, spectral analysis requires accurate modelling of synthetic stellar spectra for late-type stars, for which the assumption of local thermodynamic equilibrium (LTE) has been shown to be insufficient in many cases. Errors associated with LTE can be particularly large for \ion{Ti}{I}, which is susceptible to over-ionisation, particularly in metal-poor stars. }
  {The aims of this work are to study and quantify the 1D non-LTE effects on 
  titanium abundances across the Hertzsprung-Russell diagram for a large sample of stars.}
   {A large grid of departure coefficients, $\beta_\nu$, were computed on standard \texttt{MARCS} model atmospheres. The grid extends from $3000\,\mathrm{K}$ to $8000\,\mathrm{K}$ in $\teff$, $-0.5\,\dex$ to $+5.5\,\dex$ in $\logg$, 
   and $-5.0$ to $+1.0$ in $\feh$, with non-LTE effects in this grid reaching up to 0.4 dex.
   This was used to compute abundance
   corrections that were subsequently applied to the LTE abundances of over 70$\,$000 stars selected from the GALAH survey in addition to a smaller sample of literature Keck data for metal-poor dwarfs.}
   {The non-LTE effects grow towards lower $\feh$, lower
   $\logg$, and higher $\teff$, with a minimum and maximum $\Delta\, \mathrm{A(Ti)_{Ti\,I}}$ of $0.02$ and $0.19$ in the GALAH sample. For metal-poor giants, the non-LTE modelling reduces the average ionisation imbalance (\imbalance) from $-0.11\,\dex$ to $-0.01\,\dex$ at [Fe/H] = $-1.7$, and the enhancement in titanium abundances from
   \ion{Ti}{I} lines results in a $\tife$ versus $\feh$ trend
   that more closely resembles the behaviour of \ion{Ti}{II} at low metallicities. At higher metallicities, the results are limited by the precision of the GALAH DR3 LTE abundances and the effects are within the errors.
   For the most metal-poor dwarfs from the Keck sample, the average ionisation imbalance increases from $-0.1$ dex to $+0.2$ dex, a shortcoming that is consistent with previous 1D non-LTE studies and which we speculate could be related to 3D effects.}
   {Non-LTE effects on titanium abundances are significant. Neglecting them may alter our understanding of Galactic chemical evolution. We have made our grid of departure coefficients publicly available, with the caveat that the Ti abundances of metal-poor dwarfs need further study in 3D non-LTE.}


   \maketitle   
%

\section{Introduction}

\begin{figure*}
\centering
\includegraphics[width=2\columnwidth]{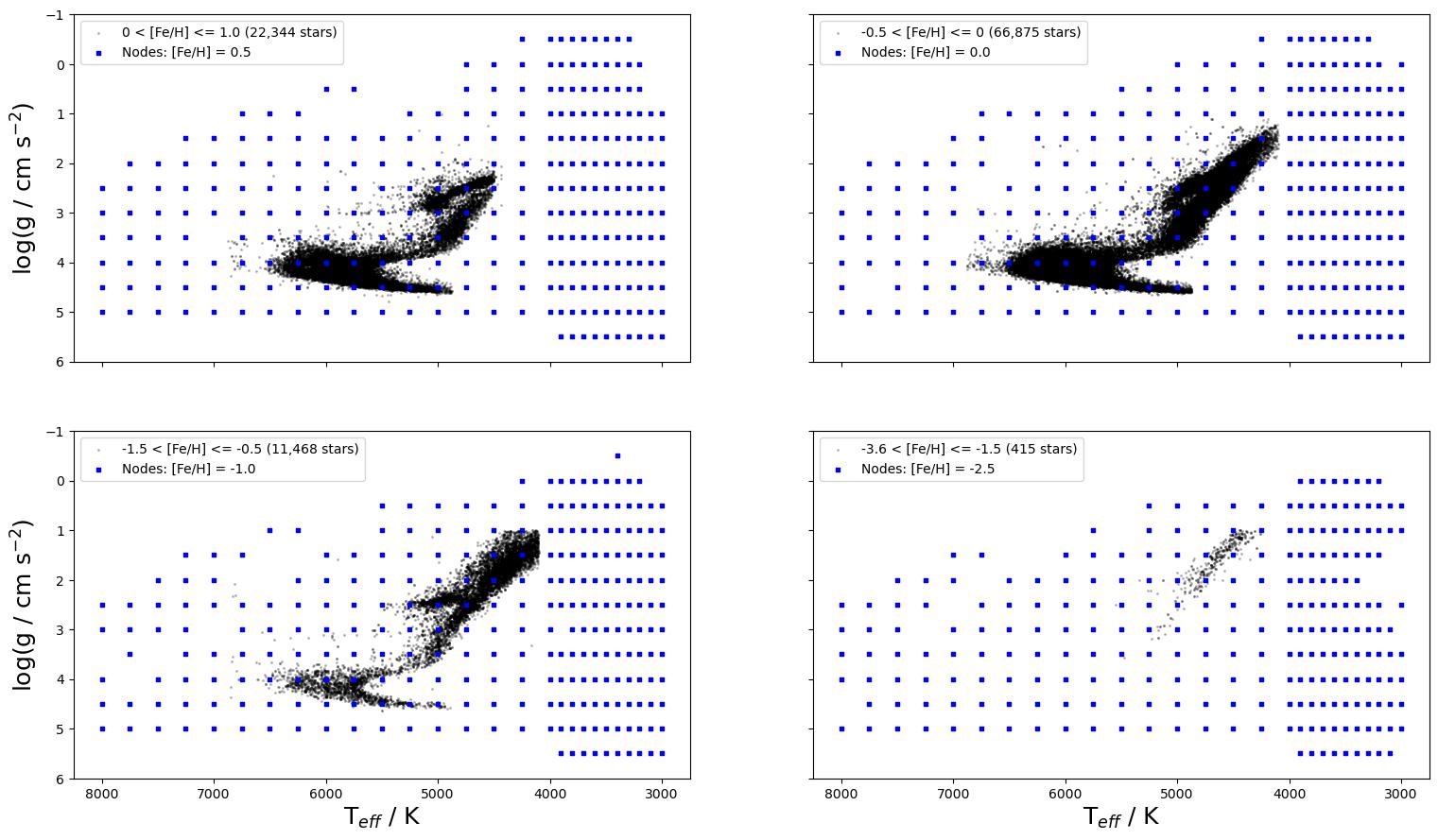}
  \caption{Kiel diagrams showing the values of $\teff$ and $\logg$ at which the non-LTE calculations were performed with \texttt{PySME}, as well as the selected GALAH stars for different metallicity ranges.}
  \label{gridnodes}
\end{figure*}

\begin{table*}
 \caption{Line list of GALAH lines.\ }

\label{tab:linelist}
    \centering
    \begin{tabular}{c c c c c c c c}

        $\lambda_{\mathrm{air}}$ (\AA) & $\abund_{\text{\sun,\, LTE}}$ & $\abund_{\text{\sun,\, NLTE}}$ & log(gf) & $E_{\mathrm{low}}/\mathrm{eV}$ &VdW& Clean \\
            \noalign{\smallskip}
        \hline\\

        \multicolumn{1}{c}{ \ion{Ti}{I} }\\
        \noalign{\smallskip}

         $4758.12$ & $4.70\pm0.05$ & $4.76\pm0.05$ & $0.51$ & $2.249$ & 326.246 &Y \\
         $4759.27$  & $4.72\pm0.01$ & $4.78\pm0.01$ & $0.59$ & $2.256$  &327.246&Y \\
         $5689.46$ & $4.83\pm0.08$ & $4.88\pm0.08$ & $-0.36$ & $2.297$ & 794.242 &Y \\
         $5716.44$ & $4.90\pm0.10$  & $4.95\pm0.10$           & $-0.72$& 2.297 & $790.243$ & Y \\
         $5739.47$ & $4.82\pm0.09$ & $4.87\pm0.09$ & $-0.61$ & $2.249$ & 327.246&Y \\

    \noalign{\smallskip}
    \hline\\

    \multicolumn{1}{c}{ \ion{Ti}{II} }\\
    \noalign{\smallskip}

         $4719.51$  & $5.12\pm0.22$ & $5.12\pm0.22$ & $-3.32$ & $2.297$ & $-7.850$&— \\
         $4798.53$  & $4.85\pm0.05$ & $4.85\pm0.05$ & $-2.66$ & $1.080$ & $-7.860$&U \\
         $4874.01$  & $4.95\pm0.06$ & $4.95\pm0.06$ & $-0.86$ & $3.095$ & $-7.830$&U \\

    \hline
    \end{tabular}
    \tablefoot{Shown are the absorption wavelength in air, the zero-point solar abundance of the line in LTE and non-LTE, the log(gf), the excitation energy, parameters for broadening due to elastic hydrogen collisions, and the quality classification from \citet{heiterblend2020}, where `Y' is clean, `U' is uncertain, `N' is not clean (blended), and `-'  was not tested. Oscillator strengths are from
    \citet{lawler} and \citet{wood} for \ion{Ti}{I} and 
    \ion{Ti}{II}, respectively. Positive broadening parameters are from ABO theory \citep{2000A&AS..142..467B}; the value before the decimal point represents the collisional cross-section $\sigma$ at $10\,\mathrm{km\,s^{-1}}$, and the value after the decimal point the velocity parameter $\alpha$. Negative values are a logarithm of full width half maximum per unit perturber number density at $10\, 000$K. Solar abundances for each line are from \citet{galahdr3} in LTE, and with the non-LTE effect added for the Sun from \fig{abundance_dep}. Isotopic splitting was accounted for when data were available in VALD \citep{vald}, (line 5739.47).}
\end{table*}

\begin{table*}
    \caption{Line list for the Keck I/HIRES sample of \citet{sneden2023}, with the same layout as \tab{tab:linelist}.}
\label{tab:snedenlines}

    \centering
    \begin{tabular}{c c c c c c c c}

        $\lambda_{\mathrm{air}}$ (\AA) & $\abund_{\text{\sun,\, LTE}}$ & $\abund_{\text{\sun,\, NLTE}}$ & log(gf)  & $E_{\text{low}}/\mathrm{eV}$ & VdW&Clean \\
            \noalign{\smallskip}
        \hline\\

        \multicolumn{1}{c}{ \ion{Ti}{I} }\\
        \noalign{\smallskip}

         $3741.06$  & $4.87$ & $4.90$ & $-0.15$ & $0.021$& 249.247 &— \\

         $4533.24$ & $4.87$ & $4.90$ & $0.54$ & $0.848$ &347.242& — \\
         $5192.97$  & $4.87$ & $4.90$ & $-0.95$ & $0.021$& 255.243 &— \\
         $5210.38$ & $4.87$ & $4.90$ & $-0.82$ & $  0.048$& 257.243&— \\

    \noalign{\smallskip}
    \hline\\

    \multicolumn{1}{c}{ \ion{Ti}{II} }\\
    \noalign{\smallskip}

         $4443.80$  & $4.92$ & $4.92$ & $-0.71$ & $1.080$& $-7.850$& — \\
         $4468.49$  & $4.92$ & $4.92$ & $-0.63$ & $1.131$& $  -7.850$& — \\
         $4501.27$  & $4.92$ & $4.92$ & $-0.77$ & $1.116$& $-7.850$& — \\
         $4571.97$  & $4.92$ & $4.92$ & $-0.31$ & $1.572$& $-7.820$& — \\

    \hline
    \end{tabular}
    \tablefoot{Lines limited to those with $\lambda>3700$\AA. Solar values adopted from \citet{Mallinson_2022}.}
\end{table*}

Data from increasingly expansive stellar spectroscopic surveys have facilitated the advancement of our understanding of galactic chemical evolution (GCE) on a large scale \citep{Jofr__2019}. The imminent influx of new data from surveys such as WEAVE \citep{Dalton18, weave} and 4MOST \citep{2019Msngr.175....3D}, which will gather of the order of $\sim 10^7$ high-resolution spectra, will set the stage for unprecedented investigations of the stellar content in our Galaxy.  

However, the impact and reliability of such studies are contingent on the accuracy of the applied spectral analysis methods and the underlying modelling of the spectral absorption lines.
Local thermodynamic equilibrium (LTE) is a common assumption in the modelling of spectral lines in late-type stars, including our Sun.  In the regime of LTE, collisional transitions dominate over radiative ones, and as such the populations of excited and ionised states can be described by the Saha and Boltzmann distributions \citep{LTEref,2015tsaa.book.....H}. This simplification offers the advantage of fast computation times, which is especially important when analysing large datasets with many spectra. However, this convenience comes at the cost of ignoring potentially important physics. 

In the case of titanium, previous 1D LTE work has demonstrated that abundances depend heavily on the diagnostic species: neutral or singly ionised. These so-called ionisation imbalances can be particularly severe in metal-poor stars \citep{metalpoor}. For example, 1D LTE analyses of the metal-poor giant HD122563 have an ionisation imbalance of \imbalance{}$= -0.4$ to $-0.45\,\dex$ \citep{Sitnova_2020,Mallinson_2022}.

Non-LTE analyses have been shown to affect and often reduce this ionisation imbalance in many cases
\citep{berge, sitnova2016, Sitnova_2020, Mallinson_2022}, with the exception of metal-poor dwarfs. 
Departures from LTE can be taken into account by solving for the statistical equilibrium \citep[e.g.][]{2003rtsa.book.....R}.  This allows external radiation to impact the local medium, which is vital for the accurate modelling of \ion{Ti}{I}; it is the minority species in late-type stars, and with an ionisation energy of $6.8\,\mathrm{eV,}$ it has a number of continua in the blue, making it sensitive to over-ionisation. These effects are enhanced in metal-poor stars due to the lack of free electrons from metals such that there are fewer electron collisions to bring \ion{Ti}{I} to LTE, and because there is less UV opacity and consequently a larger over-ionising flux.  These effects on \ion{Ti}{I} do not strongly impact the \ion{Ti}{II} lines, however, given that that \ion{Ti}{II} is typically the majority species in late-type stars; \ion{Ti}{II} also has a higher ionisation energy of $13.6\,\mathrm{eV,}$ which prohibits it from experiencing analogous over-ionisation effects.

Inelastic hydrogen collisions are typically a source of large uncertainty in non-LTE calculations, especially for metal-poor stars \citep{barklemreview2016}.  
Past studies, such as \citet{berge} and \citet{sitnova2016}, have carried out non-LTE calculations for titanium using the Drawin recipe for hydrogen collisions \citep{steenbock1984statistical,lambert_hydrogen}, while more recent studies \citep[e.g.][]{Sitnova_2020, Mallinson_2022} have improved upon this by employing the asymptotic models of \citet{belyetal2017SimplifiedModel} and \citet{belyVoronov2018Simplifiedmodel}. \citet{Mallinson_2022} also implemented a more extensive model atom and more recent data
from \citet{grumer} based on the asymptotic model of \citet{Barklem2016ca}, combined with the recipe from \citet{kaulakys} as suggested in \citet{anishOH, anishC}.

The calculations in \citet{Sitnova_2020} and \citet{Mallinson_2022} show improvements in the ionisation balance in metal-poor giants, in the latter case reducing the ionisation imbalance of HD122563 from \imbalance$= -0.45\,\mathrm{dex}$ in LTE to \imbalance$= -0.16\,\mathrm{dex}$ in non-LTE.  However, for the standard dwarf and subgiant stars considered in \citet{Mallinson_2022}, departures from LTE appeared to make the ionisation imbalance more severe.  Nevertheless, that small-scale  study of a few standard stars was not sufficient to conclusively assess the impact of non-LTE modelling across a broad range of parameter space, especially given that 1D hydrostatic model atmospheres were employed in combination with non-spectroscopic stellar parameters. 

In this work we aim to more thoroughly study and quantify the 1D non-LTE effects on titanium abundances via the analysis of a large dataset.
Following \citet{Amarsi_2020}, we constructed a grid of departure coefficients\footnote{Our grid of departure coefficients is publicly available at \url{https://zenodo.org/record/10753497}.} (\sect{method}) and used it to correct line-by-line LTE titanium abundances for $70\,347$ stars (\sect{results}) selected from the GALAH survey \citep{Martell2017, galahdr3} and 13 metal-poor stars observed with Keck \citep{sneden2023}.
We used the results to assess the impact of our non-LTE modelling on titanium ionisation imbalances as well as on our understanding of its Galactic evolution (\sect{discussion}). We present our overall conclusions in \sect{conclusion}.

\section{Method}
\label{method}


\subsection{Model atom}
\label{atom}

The model atom used in this work is based on that constructed in \citet{Mallinson_2022}, and the full details are presented therein.  The model 
contains $459$ \ion{Ti}{I} levels, $127$ \ion{Ti}{II} levels, and the ground state of \ion{Ti}{III} ($587$ levels in total).  All fine structure was collapsed. All \ion{Ti}{I} and \ion{Ti}{II} lines were radiatively coupled to the target ground states of the next ionisation stage.
In this work the model contains $5\, 026$ bound-bound radiative transitions, and the wavelength resolution of these lines was reduced such that the atom contained $63$ thousand frequency points overall, compared to the$157$ thousand in \citet{Mallinson_2022}; this had no noticeable impact on final abundances.  Inelastic collisions with free electrons were based on \citet{regemorter}  and \citet{allen}, and inelastic collisions with neutral hydrogen were based on \citet{grumer} and \citet{kaulakys} for \ion{Ti}{I}, and on \cite{Sitnova_2020} for \ion{Ti}{II}.

\subsection{Grid of departure coefficients with \texttt{Balder}}
\label{grid}

The departure coefficient, $\beta_{i}$, for a level $i$ is the ratio of populations between LTE and non-LTE:
\begin{equation}
    \beta_i=\frac{n_i}{n_i^*}
.\end{equation}
These departure coefficients for titanium were calculated in the same way as discussed in \citet{amarsi_grid2} 
for iron, using the MPI-parallelised non-LTE radiative transfer code \texttt{Balder}, which is a modified version of \texttt{Multi3D} \citep{leenaarts2009second}. The equation of state and background bound-bound, bound-free, and free-free opacities were calculated with the \texttt{Blue} package \citep{blue}.  
Details of the included species are provided in \citet{yiublue}.  In particular, Thomson scattering from electrons, and Rayleigh scattering on the red wing of Ly-$\alpha$, were included, while the opacity from background lines was treated in pure absorption.

As described in \citet{Amarsi_2020}
and \citet{amarsi_grid2},
the calculations were carried out for $3\,756$ \texttt{MARCS} model atmospheres \citep{MARCSdatabase} spanning 
$3\,000\,\mathrm{K}$ to $8\ 000\,\mathrm{K}$ in $\teff$,
$-0.5\,\mathrm{dex}$ to $5.5\,\mathrm{dex}$ in $\logg$, and
$-5.0\,\mathrm{dex}$ to $+1.0\,\mathrm{dex}$ in $\feh$.
Plane-parallel models with microturbulence $\vturb=1.0\,\kms$ 
were used for $\logg\geq4.0$,
and spherical models with solar mass and $\vturb=2.0\,\kms$ 
were used otherwise for consistency with \citet{scate}. 
For each of these grid nodes,
the calculations with \texttt{Balder} were carried out for 
different choices of titanium abundance, with
$\tife\in[-2.0,-1.0,-0.5,0.0,+0.5,+1.0,+2.0]$.  The microturbulence used in \texttt{Balder} was fixed to that used for the construction
of the model atmosphere, namely $\vturb=1.0\,\kms$ 
for $\logg\geq4.0$, and $\vturb=2.0\,\kms$ otherwise.

It should be noted that the model atom
adopts level configurations and spectroscopic terms based on those from \citet{K16}. Following \citet{Amarsi_2020}, the departure coefficients were mapped from these levels onto those given in NIST \citep{nist}.

\subsection{Abundance corrections with \texttt{PySME}}
\begin{figure*}
\centering
\includegraphics[width=8cm, height=8cm]{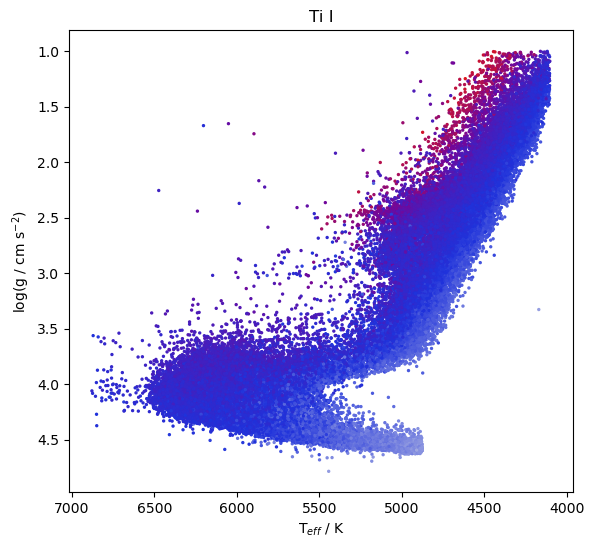}
\includegraphics[width=9.1cm,  height=8cm]{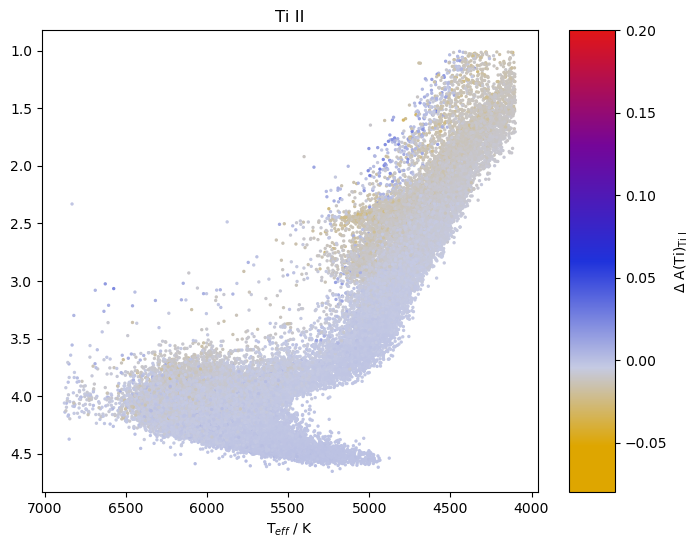}
  \caption{Non-LTE abundance correction for $\abund_{\mathrm{\ion{Ti}{I}}}$ and $\abund_{\mathrm{\ion{Ti}{II}}}$, averaged over lines, for the GALAH sample across the Kiel diagram.}
     \label{nlteimpactHR}
\end{figure*}

\begin{figure*}
\centering
\includegraphics[width=18cm]{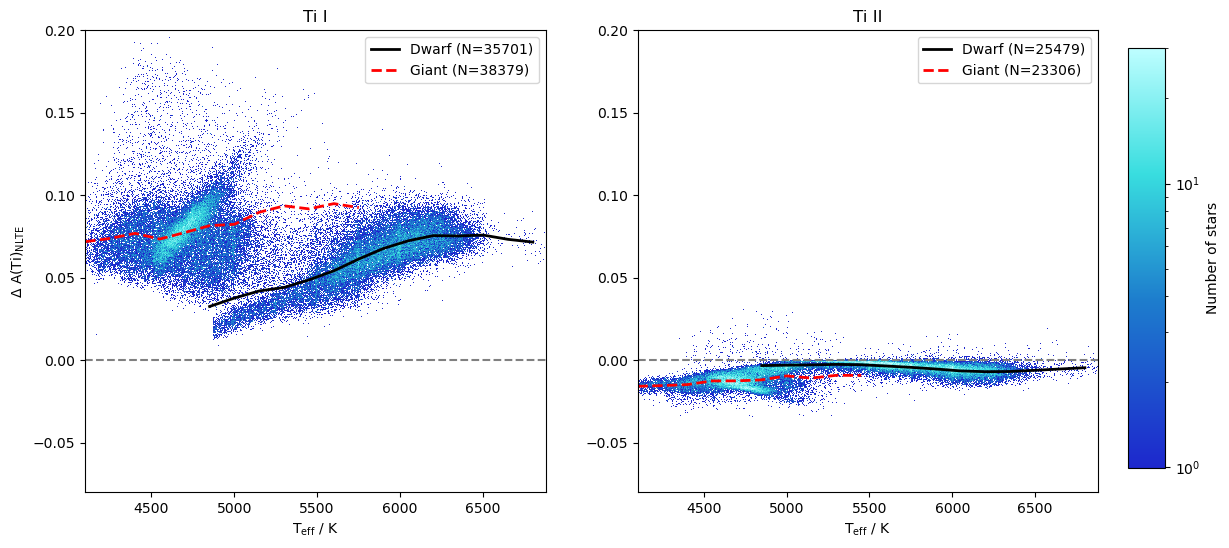}
  \caption{Non-LTE abundance correction for $\abund_{\mathrm{\ion{Ti}{I}}}$ and $\abund_{\mathrm{\ion{Ti}{II}}}$, averaged over lines, as a function of $\teff$ for the GALAH sample. The trends for dwarfs and giants are shown as solid and dashed lines, respectively, using bins of $\mathrm{\Delta\teff = \pm75K}$ and
limited to bins containing at least 30 stars. The colour gradient represents the number of stars per pixel.}
     \label{nlteimpact_teff}
\end{figure*}

\begin{figure*}
\centering
\includegraphics[width=18cm]{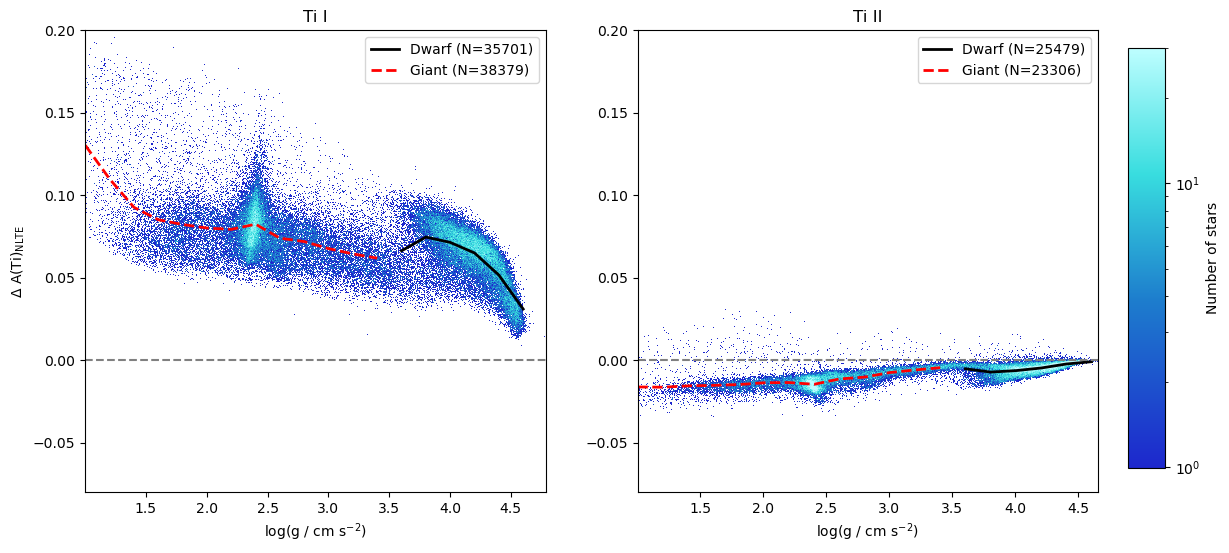}
\caption{As in \fig{nlteimpact_teff}, but with a bin size of $\Delta\logg = \pm 0.1$.}
     \label{nlteimpact_lgg}
\end{figure*}

\begin{figure*}
\centering
\includegraphics[width=18cm]{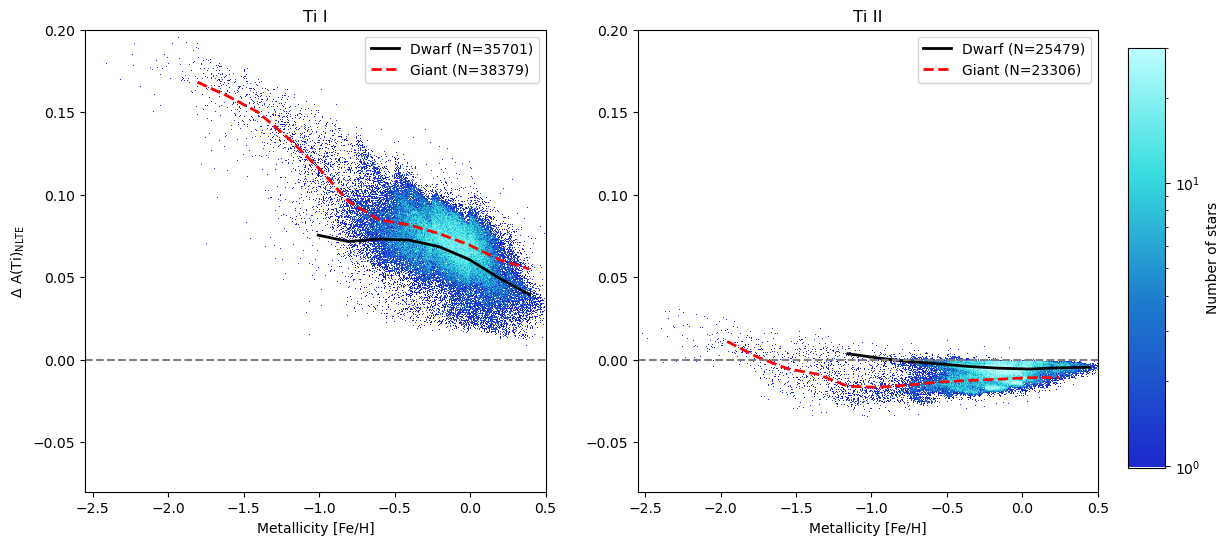}
\caption{As in \fig{nlteimpact_teff}, but with a bin size of $\Delta\feh = \pm 0.1$.}
     \label{nlteimpact_feh}
\end{figure*}

\begin{figure*}
\centering
\includegraphics[width=9cm]{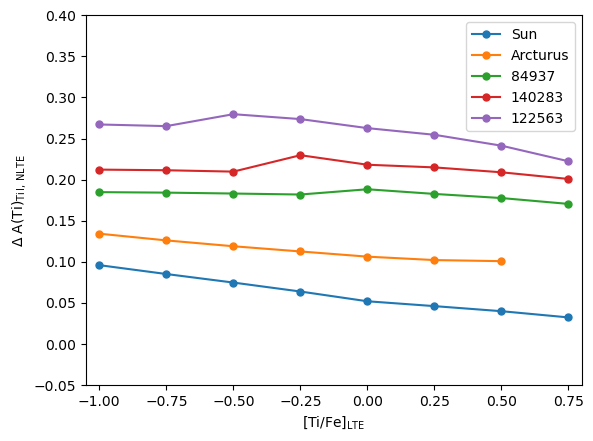}
\includegraphics[width=9cm]{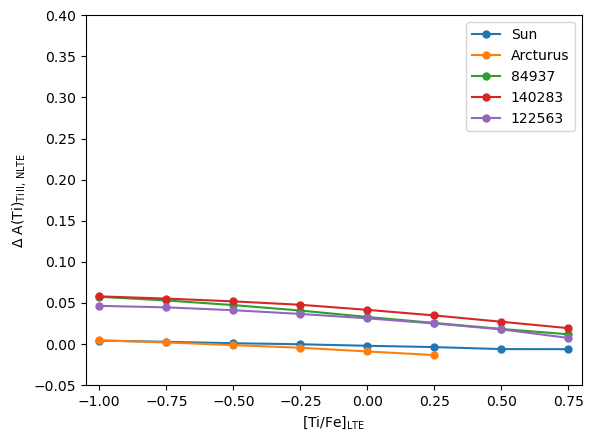}
  \caption{Trend lines showing the non-LTE effects for unsaturated lines ($W_{\mathrm{red}, \lambda} = \mathrm{log(W/\lambda)} < -4.9$) of \ion{Ti}{I} across five benchmark stars: the Sun (5773, 4.44, 0.00, 1.0), Arcturus (4286, 1.64, $-0.53$, 1.3), HD84937 (6356, 4.06, $-2.06$, 1.2), HD140283 (5792, 3.65, $-2.36$, 1.3), and HD122563 (4636, 1.40, $-2.50$, 1.8) with parameters of $\teff,\ \logg,\ \feh,$ and $\vturb$, respectively.}
     \label{abundance_dep}
\end{figure*}

\begin{figure*}
\centering
\includegraphics[width=18cm]{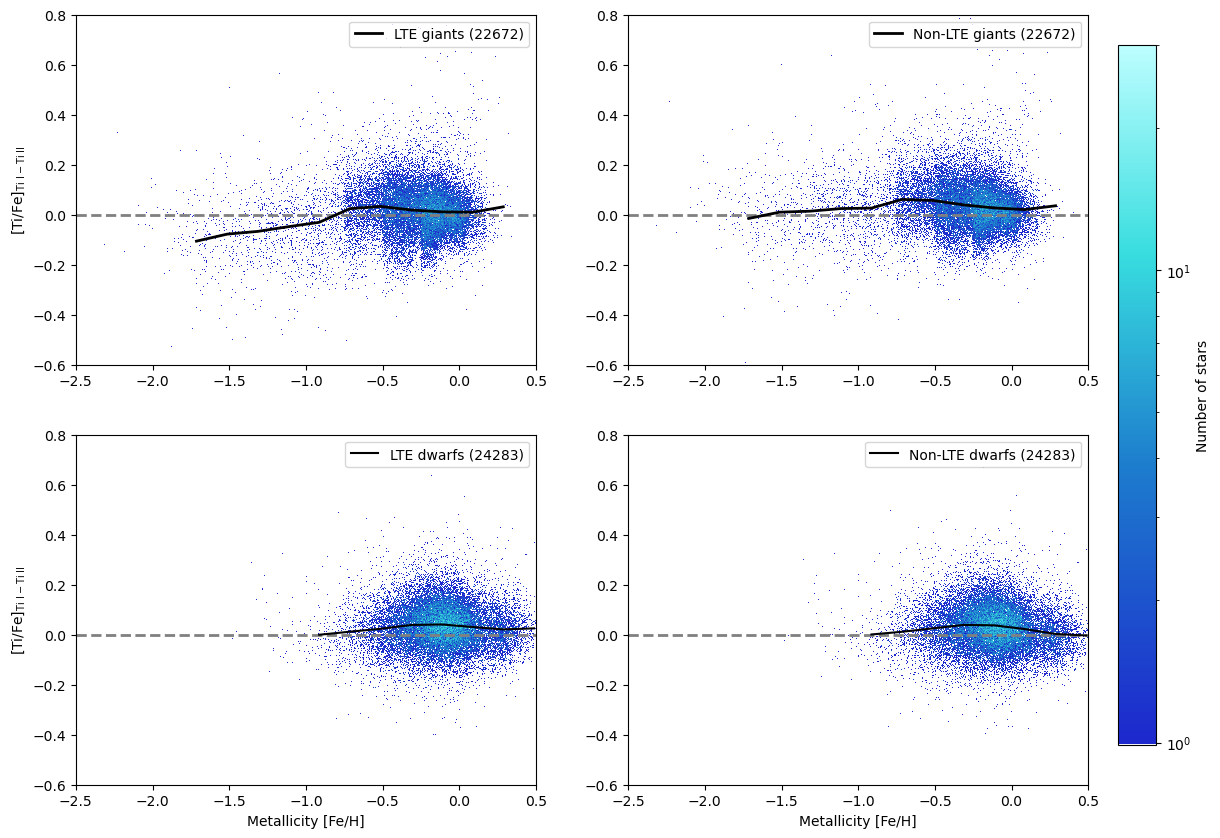}
  \caption{Ionisation imbalance, \imbalance{}, of titanium at various metallicities in LTE (left) and non-LTE (right) for giants (upper) 
  and dwarfs (lower) from the GALAH sample. The colour gradient and trend lines are the same as in \fig{nlteimpact_feh}.}
     \label{ionimb}
\end{figure*}

\begin{figure}
\centering
\includegraphics[width=9cm]{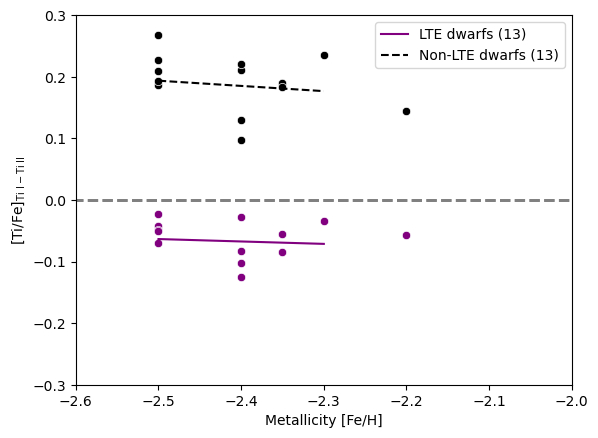}

  \caption{Ionisation imbalance, \imbalance, of titanium in metal-poor dwarfs and subgiants from the Keck sample \citep{sneden2023} in both LTE and non-LTE for the lines shown in \tab{tab:snedenlines}. Trend lines are the same as in \fig{nlteimpact_feh}.}
     \label{snedenimbalance}
\end{figure}

\begin{figure*}
\centering
\includegraphics[width=18cm]{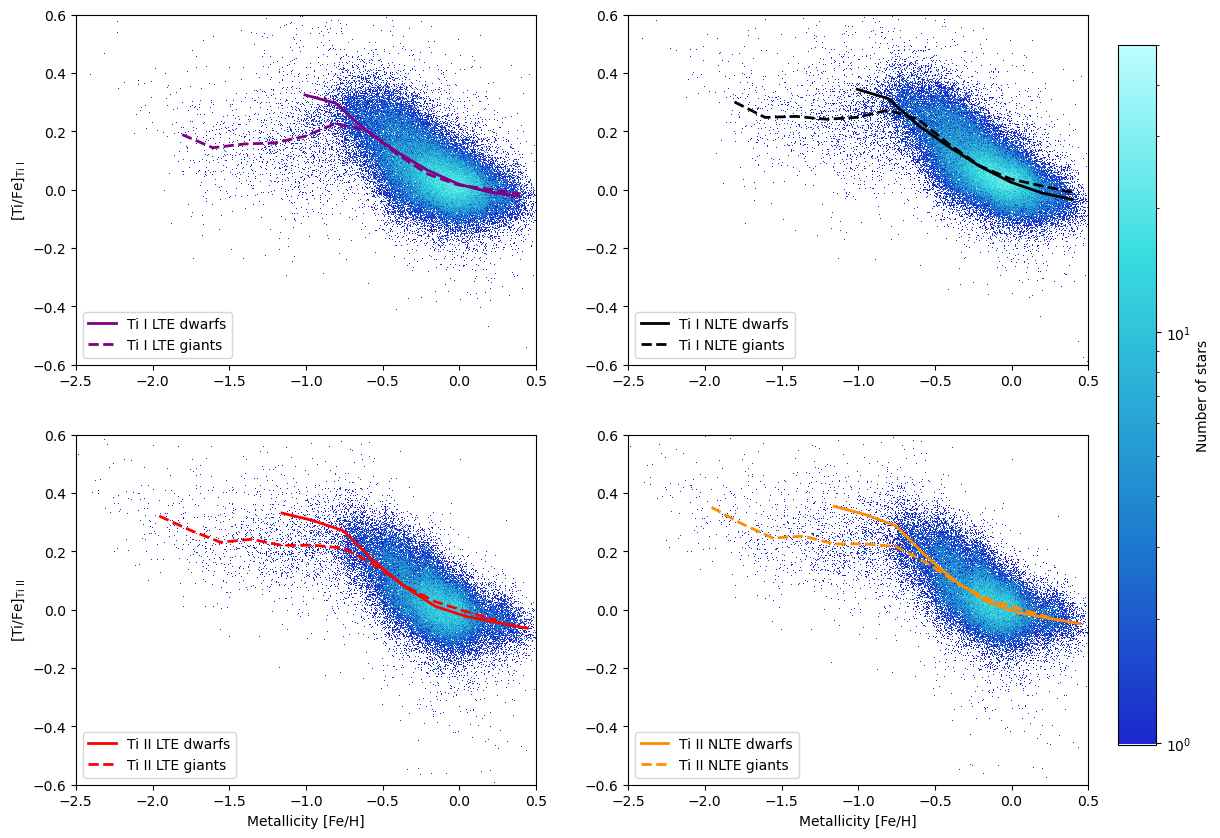}
  \caption{Run of $\tife$ versus $\feh$ in LTE (left) and non-LTE (right) from \ion{Ti}{I} (upper) and \ion{Ti}{II} (lower) for the GALAH sample. Trend lines are the same as in \fig{nlteimpact_feh}.}
     \label{trend1}
\end{figure*}

\begin{figure}
\centering
\includegraphics[width=\columnwidth]{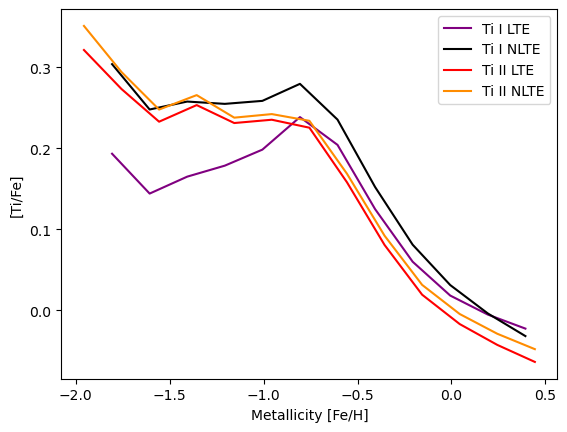}
  \caption{Trend of $\tife$ versus $\feh$ in LTE and non-LTE from \ion{Ti}{I} and \ion{Ti}{II}.
  Based on the binning from \fig{trend1} for the GALAH sample, averaged over dwarfs and giants.}
     \label{trendions}
\end{figure}

\begin{figure*}
\centering
\includegraphics[width=18cm]{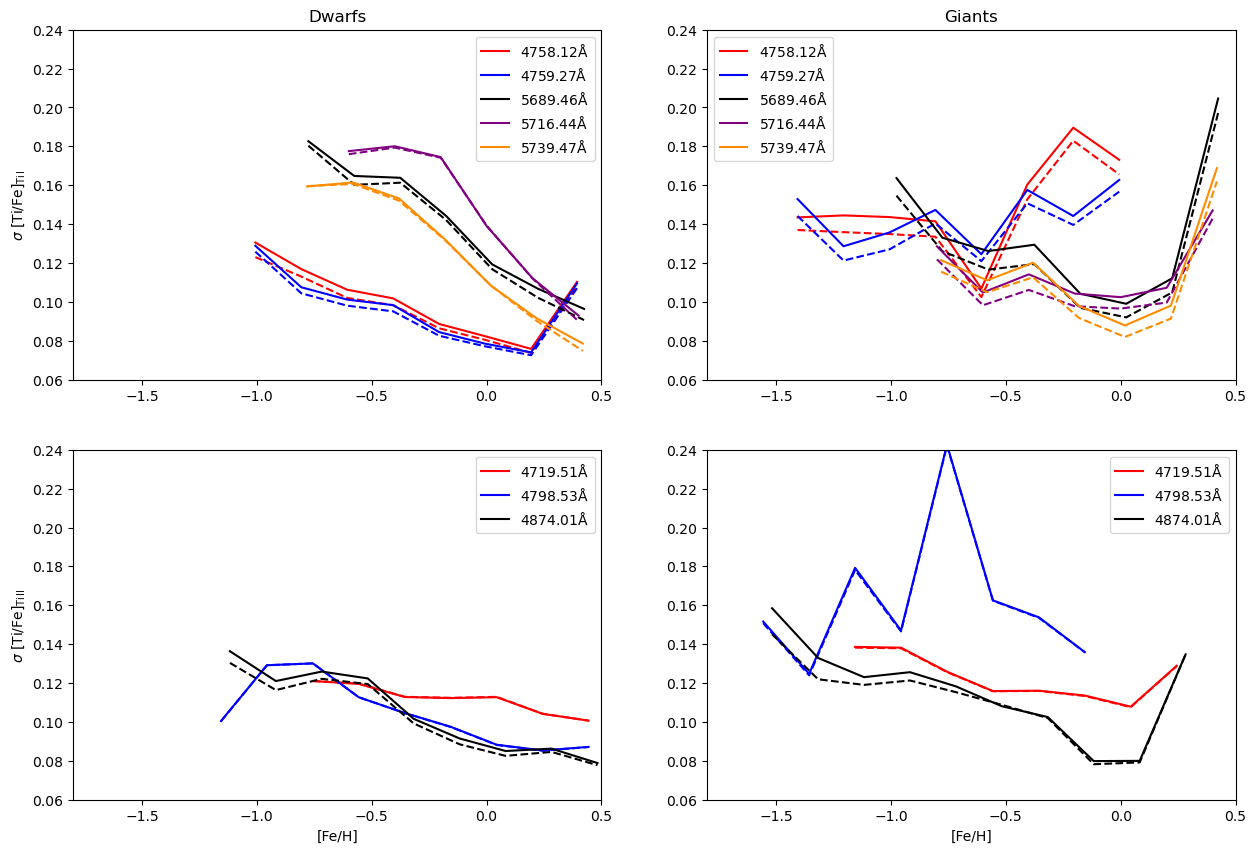}
  \caption{Star-by-star scatter for dwarfs (left) and giants (right) for different lines of \ion{Ti}{I} (upper) 
  and \ion{Ti}{II} (lower) from the GALAH sample. The colour gradient and the trend lines are the same as in \fig{nlteimpact_feh}}
     \label{starscatter}
\end{figure*}

\begin{figure*}
\centering
\includegraphics[width=18cm]{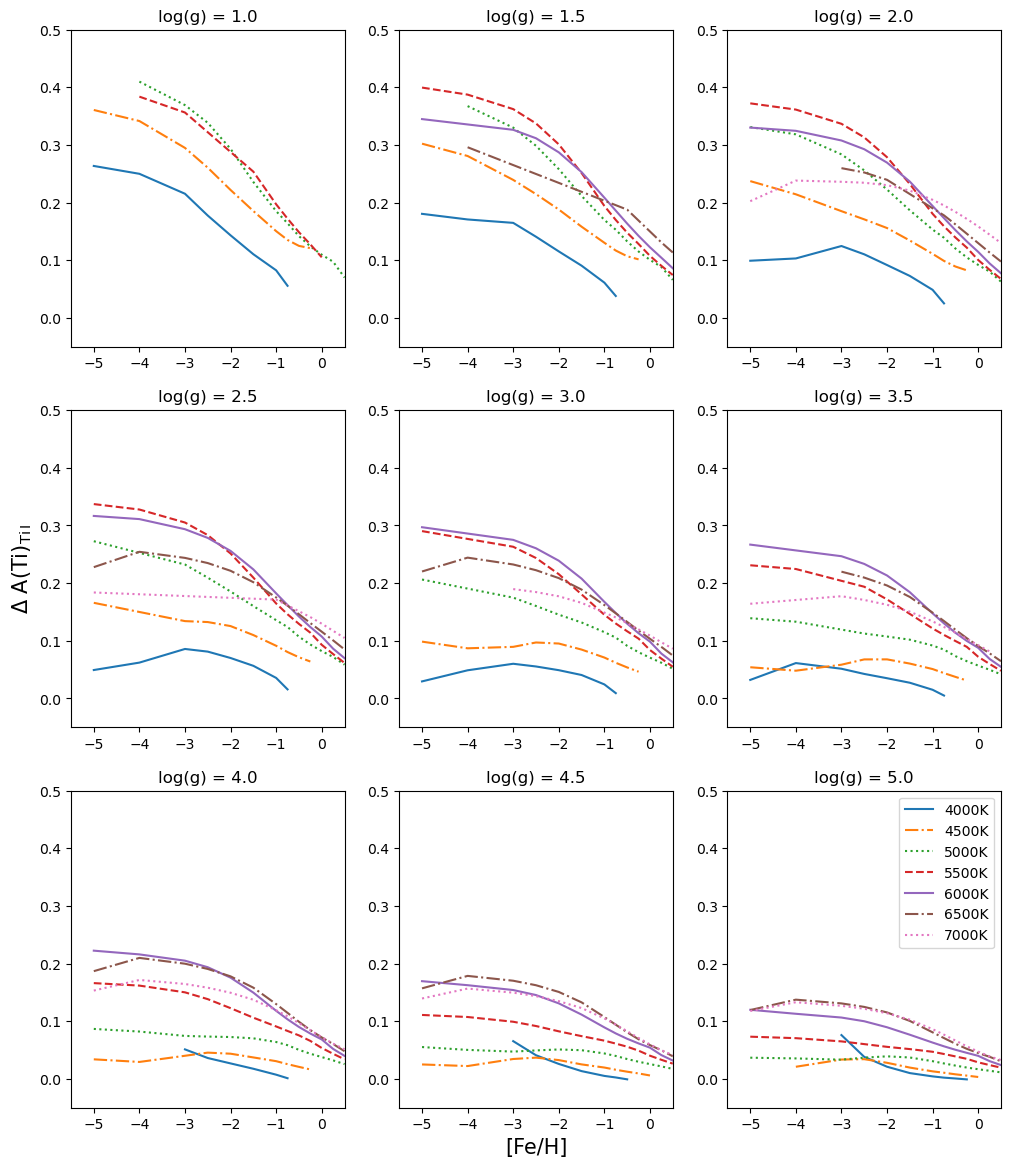}
  \caption{Trend lines showing the non-LTE effects for unsaturated lines ($W_{\mathrm{red}, \lambda} = \mathrm{log(W/\lambda)} < -4.9$) of \ion{Ti}{I} across a range of stellar parameters. All models shown have $\xi_{\mathrm{t}}$ = 2 $\mathrm{km\, s^{-1}}$ and at least two lines from Table \ref{tab:linelist}. }
     \label{nlte_spaces}
\end{figure*}

The departure coefficients from \sect{grid} were constructed in such a way as to be compatible with the spectrum synthesis codes \texttt{SME} and \texttt{PySME} \citep{Wehrhahn_2023}. For each Ti line, \texttt{PySME} searches for its upper and lower level in the grid of departure coefficients, and uses them to correct the LTE level populations before synthesising the line with the correct non-LTE line opacity and line source function \citep{nltedeparturepysme}.

As in \citet{lind2022}, \texttt{PySME} was used to generate abundance corrections using the departure coefficients for more values of $\tife$ and $\vturb$. Namely, synthetic spectra were calculated for $\tife$ between $-2.0\,\mathrm{dex}$ and $+2.0\,\mathrm{dex}$ in steps of $0.25\,\mathrm{dex}$, and $\vturb=1\,\kms$ and $2\,\kms$. These were performed on a subset of the \texttt{MARCS} model atmospheres; \fig{gridnodes} shows these grid nodes, as well as the selected GALAH stars (described in \sect{selection}). For a given titanium line, equivalent widths were calculated in LTE and non-LTE and linearly interpolated to generate a grid of abundance corrections shown in \fig{nlte_spaces}.  Finally, these abundance corrections were linearly interpolated onto the line-by-line LTE GALAH abundances, as functions of $\teff$, $\logg$, $\feh$, $\vturb$, and abundance A$(\mathrm{Ti})_{\text{line, \,LTE}}$.  The star selection is described in \sect{selection} and the line selection is described in \sect{linelist}, below.


\subsection{GALAH DR3 star selection}
\label{selection} 

This work builds upon the few benchmark stars analysed in \citet{Mallinson_2022} and examines a subsample of $70\,347$ stars, all of which have some measurement of titanium abundance from
either \ion{Ti}{I} or \ion{Ti}{II} (or both), out of the $588\,571$ stars in GALAH DR3 \citep{galahdr3}.
These data contain $\teff$, $\logg$, and $\feh$ (both the initial atmospheric and final estimate of abundance, with the latter being used in this work), $\vturb$, and line-by-line LTE titanium abundances, $\abund_{\text{line,\,LTE}}$. The way in which these parameters were determined are described in \citet{galahdr3}.  To give a short summary, these parameters were determined by normalising the spectra based on some initial guess, and optimising the stellar parameters. Forming an initial $\logg$ by the mass-luminosity relation, $\teff$, $\feh$, $v_{\mathrm{broad}}$, and $v_{\mathrm{rad}}$ are optimised with double-sided partial derivatives and $\chi^2$, updating $\logg$ and $v_{\mathrm{mic}}$ with each change, repeating this until the change in $\chi^2$ was below 0.001 for each iteration. [Fe/H] was finally redetermined in non-LTE via analysing Fe lines directly (and this value of [Fe/H] is adopted for the present study).

The selected stars are shown in \fig{gridnodes}.
The selection was made by limiting to those stars with the most reliable parameters and highest signal-to-noise ratio.  This was done in several steps.  First, only the stars with quality flags of 0 for $\feh$ and $\tife$ (`flag\_fe\_h', `flag\_ti\_fe', and `flag\_ti2\_fe') and the stellar parameter flag `flag\_sp' were selected. Secondly, as recommended in the best practices for using the data from the GALAH consortium\footnote{\url{https://www.galah-survey.org}}, the average signal-to-noise ratio per pixel for the third charged couple device was limited to $30$ or above ($\text{snr\_c3\_iraf} > 30$) for very metal-poor stars ($\feh < -1.5$), in order to keep enough low metallicity stars in the sample for a robust analysis, and average signal-to-noise ratio over all lines of $\mathrm{S/N} > 70$ for the stars at higher metallicities. Finally, only stars with at least two lines with a reduced equivalent width $\mathrm{W_{red, \lambda}} = \mathrm{log(W}_{\lambda}/ \lambda)$ in LTE between $-6.0 < \mathrm{W_{red,\, \lambda}} < -4.9$ were kept.


\subsection{Line selection}
\label{linelist} 

As the GALAH reanalysis of absolute abundances, $\abund{}$, was done on a line-by-line basis, the line-by-line scatter can be used as another diagnostic of the non-LTE modelling.  
The line selection strongly influences the final results, and care was taken to try to select the best available lines out of those considered by GALAH.

\ion{Ti}{I} lines were initially limited to those that are reported as unblended in \citet{heiterblend2020} in the Sun and Arcturus, although the 6599.11\AA$ $ line was also removed due to a lack of titanium abundance data in the available stars.

In the case of \ion{Ti}{II}, no lines were reported as unblended, so lines were instead restricted to those that were reported as uncertain.
Of the remaining available lines, 4865.6$\AA$ can be strongly affected by H$\beta$, causing blending and continuum normalisation, primarily in dwarfs. However, to keep the line list identical for all stars, it was removed entirely. To increase the line list to more than two lines, the least visibly blended line in the Sun and Arcturus that was not examined in \citet{heiterblend2020}, at 4719.51$\AA$, was added.

The relative abundances $\tih$ and $\tife$ were determined on a line-by-line basis using the GALAH abundances, with non-LTE solar corrections applied to their values when in non-LTE. Average values of $\tih$ were calculated for different species, $\tih_{\mathrm{\ion{Ti}{I}}}$ and $\tih_{\mathrm{\ion{Ti}{II}}}$,
via a mean of the available lines for a given star weighted by their error. Weak lines and strong lines were cut using the reduced equivalent
width criterion of $-6.0 < \mathrm{W_{red,\, \lambda}} < -4.9$. The upper limit prevents the use of saturated lines, which would have both given an under-abundance and increase the sensitivity to microturbulence error. The lower limit reduces the impact of noise fluctuation.
The un-normalised weights were given by $w_{\text{line}}=1/\sigma_{\text{line}}^2$, using the uncertainties provided by GALAH. 
Finally, where illustrated in \sect{results} below, the absolute abundances use the non-LTE line-averaged solar correction: $0.052\,\dex$ for \ion{Ti}{I} and $-0.002\,\dex$ for \ion{Ti}{II}.


\section{Results and discussion}
\label{discussion}

\subsection{Non-LTE effects across the Hertzsprung-Russell diagram}
\label{results}

Figures \ref{nlteimpactHR}--\ref{nlteimpact_feh} show the line-averaged non-LTE corrections of \ion{Ti}{I} and
\ion{Ti}{II} as a function of stellar parameters.
The non-LTE effects on both species grow in severity towards lower metallicities, lower surface gravities, and higher effective temperatures, qualitatively similar to that found for iron in \citet{karin}.

For \ion{Ti}{I}, the abundance corrections rise to $+0.18\,\dex$ on average for giants at $\feh=-2.0$ (\fig{nlteimpact_feh}).
The analysis of benchmark stars in \citet{Mallinson_2022} suggests that the dwarfs should show a similar trend with $\feh$ as the giants, namely increasing non-LTE abundance corrections with decreasing metallicity.  However, the dwarfs in this sample only extend down to  $\feh=-1.0$,
reflecting the low number of very metal-poor dwarfs in the dataset of \citet{galahdr3}.  At $\feh=-1.0$,
the abundance corrections reach $+0.08\,\dex$, compared to $+0.13\,\dex$ for giants at this same metallicity.

The abundance corrections for \ion{Ti}{II} are generally much less severe than for \ion{Ti}{I}.
This is because of the majority status of \ion{Ti}{II}: due to the larger number of ions, the effect of over-ionisation by UV radiation is smaller overall, which implies that the assumption of LTE for both dwarfs and giants is a more reasonable one for this species.
The corrections are typically $-0.02$ dex for giants but rising at low metallicities, and consistent with zero for the dwarfs in the GALAH sample. 
However, at the lowest metallicities around $\feh=-2.0$ the abundance corrections reverse sign and tend to become slightly positive (\fig{nlteimpact_feh}),
consistent with \citet{Mallinson_2022}. It reflects how over-excitation due to radiative pumping becomes increasingly important for this species towards lower metallicities \citep{Sitnova_2020}.

Additionally, the impact of titanium abundance itself on the non-LTE corrections for \ion{Ti}{I} was investigated in \fig{abundance_dep} for benchmark stars from \citet{Mallinson_2022}. Between $-1.00$ < [Ti/Fe] < $0.75$ the correction changes by only 0.05 dex for the Sun, and 0.09 dex for HD122563. Due to the large difference in abundance, this is not a significant change and within the error of this paper.

\subsection{Ionisation imbalances}
\subsubsection{GALAH stars}
\label{discussion_imbalance}

The difference in the predictions based on \ion{Ti}{I} and \ion{Ti}{II} is a long-standing problem in
studies of titanium abundances, with \ion{Ti}{I} predicting lower abundances than \ion{Ti}{II}.
Neglected non-LTE effects are a contributing factor to this problem  \citep{berge}. 
For both dwarfs and giants, the abundance corrections become larger towards lower metallicities for \ion{Ti}{I},
while remaining rather close to zero for \ion{Ti}{II}.

Consequently, the ionisation imbalance is reduced in non-LTE in giants from the GALAH sample at low metallicities, as shown in \fig{ionimb}. 
For giants in LTE, the ionisation imbalance reaches on average \imbalance$=-0.11,\dex$ at $\feh \approx -1.7$. 
In non-LTE, this is reduced to \imbalance$=-0.01$, demonstrating a non-LTE impact on the imbalance of $0.10$ dex. 

For dwarfs in the GALAH sample, departures from LTE do not significantly alter the ionisation imbalance on average, with an imbalance of \imbalance$=-0.000$ for LTE and \imbalance$=0.001$ in non-LTE, at $\feh \approx -0.9$.  
It should be noted that this results in a solar abundance correction of $0.05\,\dex$ for \ion{Ti}{I}
and $-0.002\,\dex$ for \ion{Ti}{II} (see \tab{tab:linelist} and \sect{linelist});
the square bracket notation thus hides an ionisation imbalance of around $0.05\,\dex$.


\subsubsection{Keck stars}

Due to the lack of low metallicity dwarfs available in the GALAH data sample, an additional set of stars were taken from \citet{sneden2023}, consisting of 13 dwarfs of the original 37 due to line selection and saturation; the majority of which have a metallicity between $-3.0 < \mathrm{\feh} < -2.0,$ as shown in \fig{snedenimbalance}. The spectra of these stars were obtained with Keck I/HIRES spectrograph \citep{vogt}, and originally used in \citet{boe}. The lines available for these stars do not overlap with those in the GALAH data, and the majority instead are below $4000\AA$. The inclusion of these lines dramatically reduces the abundance of both species of titanium.

In HD84937, \citet{lawler} found a strong reduction in \ion{Ti}{I} abundances of around $0.15-0.2$ dex when moving to the near-UV range of $\lambda = 3100-3700\AA$. \citet{wood} finds similarly for the high excitation energy lines. The overabundance of UV lines in the \ion{Ti}{II} line list may result in the heavily positive LTE ionisation imbalance \imbalance seen in \citet{sneden2023} that is in disagreement with the expectations from previous work that focused on the longer optical wavelength lines \citep{berge, Sitnova_2020, Mallinson_2022}. The cause of this may be related to the Balmer continuum opacity \citep{wood}. In \citet{2016sneden}, the removal of these lines raises the \ion{Ti}{II}-based abundance and results in a smaller imbalance, similar to that found for HD84937 in \citet{Mallinson_2022}.

By including only unsaturated and reasonably strong lines ($-6.0 < \mathrm{W_{red,\, \lambda}} < -4.8$) above this limit (i.e. $>3700\AA,$ as shown in Table \ref{tab:snedenlines}), only 13 of the original 34 stars in \citet{sneden2023} can be analysed. However, this gives the ionisation imbalance for dwarfs at these metallicities a closer agreement with \citet{Mallinson_2022} and remains close to 0 in LTE, worsening in non-LTE and prompting 3D analysis to further investigate.

\subsubsection{3D impact}

The inclusion of 3D effects is a potential solution to the ionisation imbalance of Ti. To estimate its impact in LTE, we employed four snapshots of two 3D \textsc{STAGGER} radiation-hydrodynamic models \citep{Magic_2013}. The models pertain to metal-poor dwarfs with $\teff$ = 6000K, $\logg$ = 4.5, [Fe/H] = $-3.0$ and [Fe/H] = $-2.0$, respectively.  [Ti/Fe] was set to 0.4, and we adopted $\xi = 1$\,km s$^{-1}$ in the corresponding 1D calculations. The spectral lines listed in Table 2 were computed both in 1D and 3D under LTE conditions using the \textsc{SCATE} \citep{scateog} code. The mean 3D effects (i.e. 3D minus 1D abundance differences) were found to be $-0.41$ dex and 0.01 dex for \ion{Ti}{I} and \ion{Ti}{II}, respectively, in the [Fe/H] = $-3$ model, and $-0.36$ dex and $-0.05$ dex, respectively, in the [Fe/H] = $-2$ model. 

For giants characterised by $\teff$ = 4000K, $\logg$ = 2.0, and [Fe/H] = $-2.0$, a considerably lesser impact of $-0.04$ and +0.06 is found for \ion{Ti}{I} and \ion{Ti}{II}, respectively. These findings align with prior research \citep{3Dcomplex} that reports a large abundance change in 3D for resonant lines only at around $-0.45$ dex at [Fe/H] = $-3$, but that steeply decreases to -0.05 dex in abundance change as the excitation energy increases to 2eV.

These results underscore the significant influence of 3D effects on \ion{Ti}{I} in the inverse direction of non-LTE, offering a potential resolution to the dwarf metallicity discrepancy at lower metallicities. However, the interaction between 3D and non-LTE is not necessarily additive \citep{Lagae_2023} and a full 3D non-LTE investigation would be needed to make strong conclusions.

\subsection{Star-by-star scatter}

The abundance results in \fig{trend1} also suggest a slight reduction in the
star-by-star scatter in the $\tife$ versus $\feh$ plane when departures from LTE are taken into account.
This is also illustrated in \fig{starscatter}, where we plot the standard deviation 
in the $\tife$ versus $\feh$ plane in different $\feh$ bins for given titanium lines.
There is indeed a slight reduction in the star-by-star scatter, reaching over
$0.01\,\dex$ for the \ion{Ti}{I} $4759.27\AA$ line in low-metallicity giants.
As a result, the offset between dwarfs and giants is slightly reduced
when comparing the upper left and upper right panels of \fig{trend1}.

It is plausible that the scatter of these results is dominated by issues with the LTE GALAH abundances.
As such, the non-LTE model presented here may lead to a further reduction in
the star-by-star and line-by-line scatter when applied to more precise measurements, but no strong conclusion can be made from such small effects.  


\subsection{Departures from LTE for different stellar parameters}

The impact that stellar parameters have on the magnitude of the non-LTE effect is shown in \fig{nlte_spaces}, computed using the \texttt{MARCS} atmospheres at specific grid points with \texttt(PySME), and plotted as the average non-LTE impact on A(Ti), averaged over the chosen GALAH \ion{Ti}{I} lines.
As is found in previous work, a decrease in metallicity increases the magnitude of $\Delta$ A(Ti)$_{\mathrm{Ti\, I}}$ in most cases, particularly between $-3 < \feh < 0$. 
Similarly, a decreasing $\logg$ results in a larger non-LTE effect as collisional rates — scaling with the number density of both titanium atoms and the ionising impact species — reduce faster than radiative rates, which scales only with the number density of titanium. 
As seen in \fig{nlte_spaces}, this effect also increases with $\teff$ until around $\teff \approx 6000$K, as the increased number of photons outweigh the increased number and speed of colliders. Beyond this, as with most titanium lines in \citet{sitnova2016}, the increasing temperature begins to reduce the non-LTE effect up to our limit of 7000K. The exact temperature at which this decline occurs varies with $\logg$ and $\feh$. Of the available grid points, the most extreme non-LTE effects range from -0.004 (at $\teff = 4000$K, $\logg = 2.5$, [Fe/H] $= 0.5$) to 0.442 dex(at $\teff = 5000K$, $\logg = 1.0$, [Fe/H] $=-4.0$).

\subsection{The Galactic evolution of titanium}
\label{discussion_gce}

Precision measurements of chemical abundances are crucial to accurately interpreting and constraining theories of GCE. While titanium is formally considered an iron peak element, it often behaves as an alpha element in observations \citep{Zhao2016, kobayashi2020origin} Thus, it is useful in disentangling Galactic components and tracing back their evolution histories. For example, [Ti/Fe] can be used as a chemical tracer to separate the chemical thin and thick discs, stellar halo and accreted stellar populations \citep[e.g.][]{Bensby2005, bensby, Mashonkina2019}. In addition, the Galactic Anticenter Stellar Structure, or `Monoceros Ring', has been shown to exhibit abundance patterns in [Ti/Fe] and [Y/Fe] that are similar to those of Sagittarius stars and distinct from Milky Way stars, allowing for constraints to be placed on the origin of this structure \citep{Chou2010}.

Interestingly, some recent GCE models show that [Ti/Fe] is under-predicted by $\sim 0.4$ dex over a wide metallicity range when compared to observational data \citep{Kobayashi2006, Romano2010, kobayashi2020origin}, even in non-LTE \citep{Mashonkina2019}. 
It has been shown that [Ti/Fe] levels can be increased by including yields from asymmetric 2D models of high-entropy, jet-like ejecta from hypernovae \citep{Maeda2003, Tominaga2009} in these GCE models \citep[see model K15, as presented in][]{2016sneden, kobayashi2020origin}, but high precision measurements of [Ti/Fe] are needed to constrain the parameters of these ejecting supernovae. Notably, the non-LTE corrections presented in this work have shown that average giant [Ti/Fe] abundances measured with \ion{Ti}{I} become larger by $\sim 0.1$ dex as compared to 1D-LTE analyses, thereby exacerbating the discrepancy between models and observations, particularly at low metallicities of [Fe/H] < $-0.7$. 

In a study designed to investigate the impact of modelling assumptions on GCE yields, \citet{Cote2017} emphasise the significant influence of the choice of stellar yields on the trends in chemical evolution. With different stellar models within their set of stellar yields, they substantially reduce the titanium underprediction of \citet{Cote2016} and \citet{kobayashi2020origin} and successfully replicate an approximately +0.25 enhancement in [Ti/Fe] at low metallicities, within the framework of a one-zone model, and consistent with observations in the Sculptor dwarf galaxy.
However, they point out that these abundance trends can be realised through various combinations of input parameters, making it challenging to distinguish among several degenerate models, a caveat that was also described in previous work \citep{Romano2013}. 

Clearly, the abundance trends from these models are highly dependent on the choice of input parameters and the adopted stellar yields. This implies a fundamental limitation in one-zone GCE models, which cannot be overcome by simply increasing the precision of observed abundance data.

Therefore, while the non-LTE corrections reduce the discrepancy between abundances measured using \ion{Ti}{I} and \ion{Ti}{II} for giants at low metallicities, they are not sufficient to offer further insight needed to constrain GCE models.\ They nevertheless emphasise the need for improved precision in theoretical yields for these models.

\section{Conclusion}
\label{conclusion}

This study is the first to investigate the impact of 1D non-LTE effects on titanium abundances across a large sample of stars, building upon previous work that had primarily examined a small number of benchmark stars. It was expected that larger non-LTE effects would be found for Ti I, in metal-poor stars, and in giants: all three of these predictions are confirmed in this investigation. This resulted in a reduced ionisation imbalance for the metal-poor giants, a slight reduction in the disagreement between dwarf and giant titanium abundances at lower metallicities, and a shift in the [Ti/Fe] trend of \ion{Ti}{I} to more closely match that of \ion{Ti}{II} at low metallicities. This takes observations further away from predictions by some GCE models of titanium \citep[][and references therein]{kobayashi2020origin}. Overall, this suggests that the inclusion of the departure coefficient grid and correcting for non-LTE effects results in more accurate \ion{Ti}{I} abundances for giants at lower metallicities, although this is dependent on the line selection.

Due to the small sample of metal-poor dwarfs and the relatively low accuracy of the GALAH sample, the limitations of this analysis are apparent. In \citet{Mallinson_2022}, metal-poor dwarfs were seen to have their imbalance worsened by the inclusion of non-LTE, a conclusion that is reinforced by our analysis of the \citet{sneden2023} sample in this work. An investigation of 3D non-LTE effects for Ti in metal-poor stars may shed further light on the problem. Additionally, the low number of lines and their quality — particularly for \ion{Ti}{II} — further limit the reliability of these results.

Finally, there were larger departures from LTE towards higher effective temperatures and lower surface gravities, as well as lower metallicities. However, at lower surface gravities of $\logg < 3.5$, increasing the temperature beyond $\teff \approx 5500$K reduced the departures from LTE. Likewise, a lower metallicity can occasionally result in lower non-LTE effects in all lines analysed, primarily at low temperatures and high surface gravity. In summary, we find the non-LTE corrections reach up to $+0.41$ dex at $\teff = 5000$K, $\logg = 1.0$, [Fe/H] $=-4.0$.


\begin{acknowledgements}
AMA acknowledges support from the Swedish Research Council (VR 2020-03940). This research was supported by computational resources provided by the Australian Government through the National Computational Infrastructure (NCI) under the National Computational Merit Allocation Scheme and the ANU Merit Allocation Scheme (project y89).
This work has made use of the VALD database, operated at Uppsala University, the Institute of Astronomy RAS in Moscow, and the University of Vienna.
We thank the PDC Center for High Performance Computing, KTH Royal Institute of Technology, Sweden, for providing access to computational resources and support.
JWEM, KL, and KY acknowledge funds from the European Research Council (ERC) under the European Union’s Horizon 2020 research and innovation programme (Grant agreement No. 852977) and funds from Knut and Alice Wallenberg Foundation.
We acknowledge Chris Sneden for the sharing of his data and advice in its analysis.
\end{acknowledgements}
\bibliographystyle{aa_url} 
\bibliography{main.bib}
\begin{appendix}

\section{Additional non-LTE effects and microturbulence sensitivity}
\label{appx}
\begin{figure*}
\centering
\includegraphics[width=18cm]{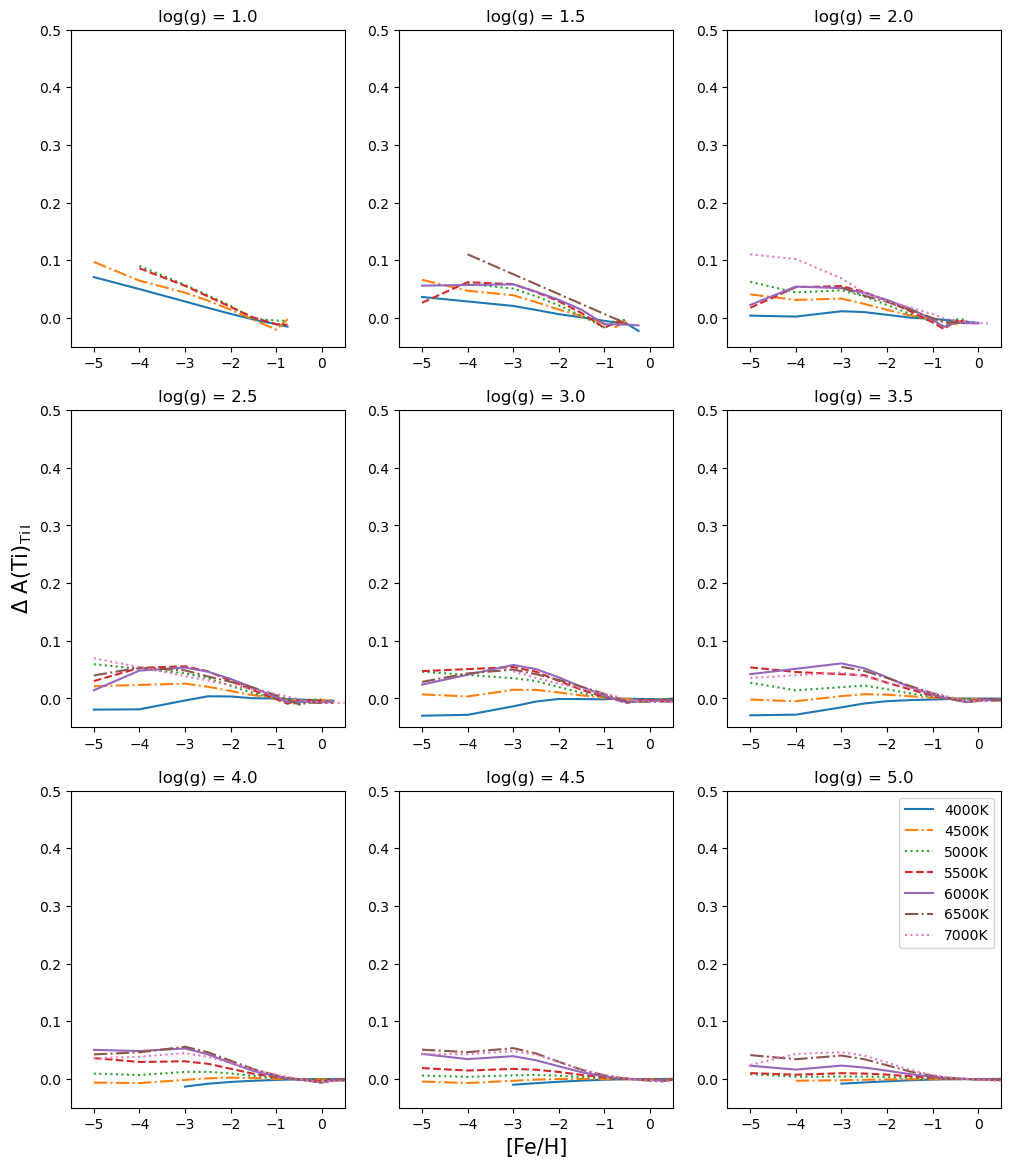}
  \caption{Trend lines showing the non-LTE effects for unsaturated lines ($-6.0 < W_{\mathrm{red}, \lambda} = \mathrm{log(W/\lambda)} < -4.9$) of \ion{Ti}{II} across a range of stellar parameters. All models shown have $\xi_{\mathrm{t}}$ = 2 $\mathrm{km\, s^{-1}}$ and at least two lines from Table \ref{tab:linelist}. }
     \label{nlte_spacesTi2}
\end{figure*}

\begin{figure*}
  \includegraphics[width=\textwidth]{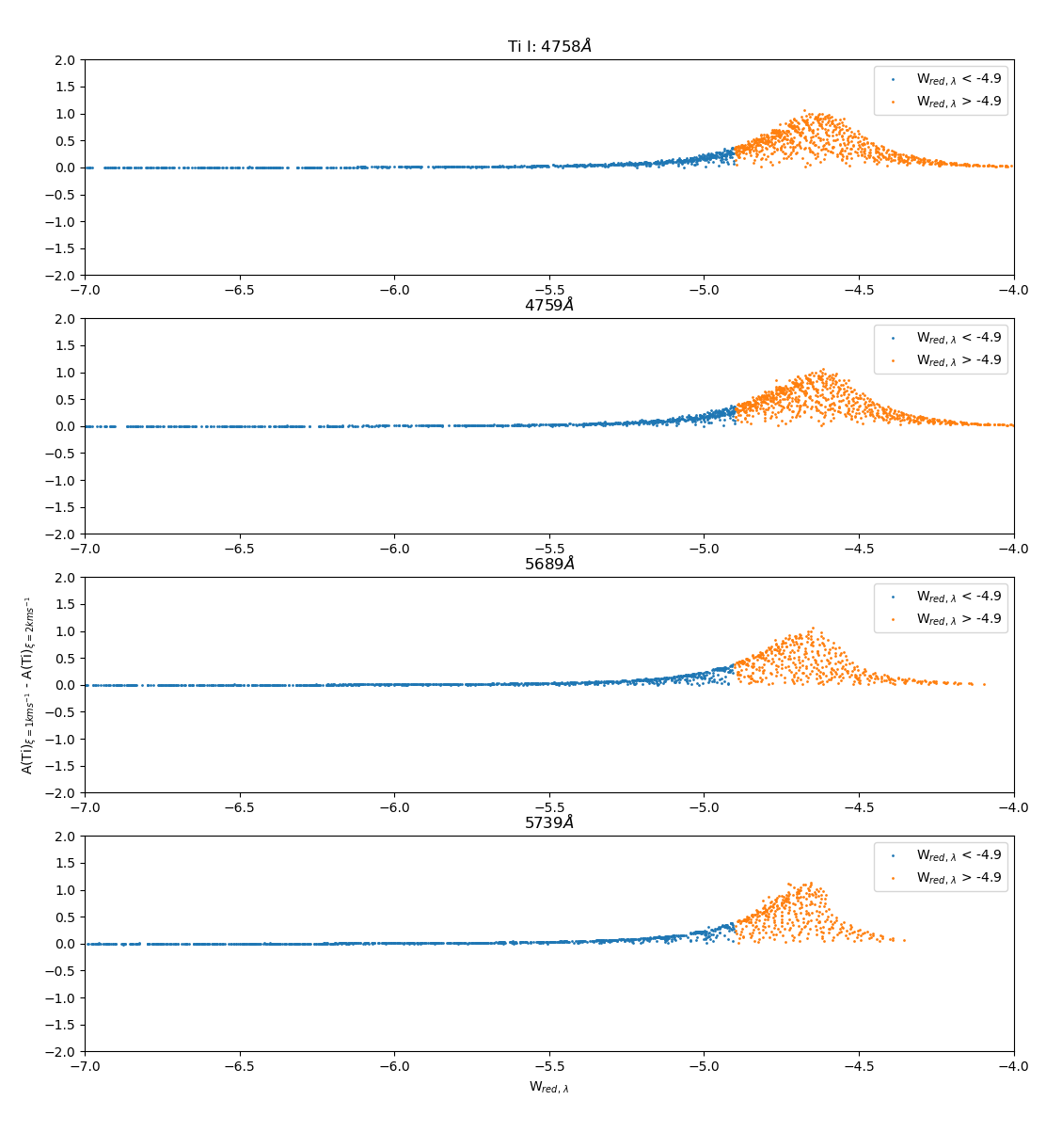}
  \caption{Difference in abundance for GALAH stars with microturbulences replaced by $\xi = 1$ and $\xi = 2$ kms$^{-1}$ for Ti I.} 
      \label{fig:vmictest}
\end{figure*}

\begin{figure*}
  \includegraphics[width=\textwidth]{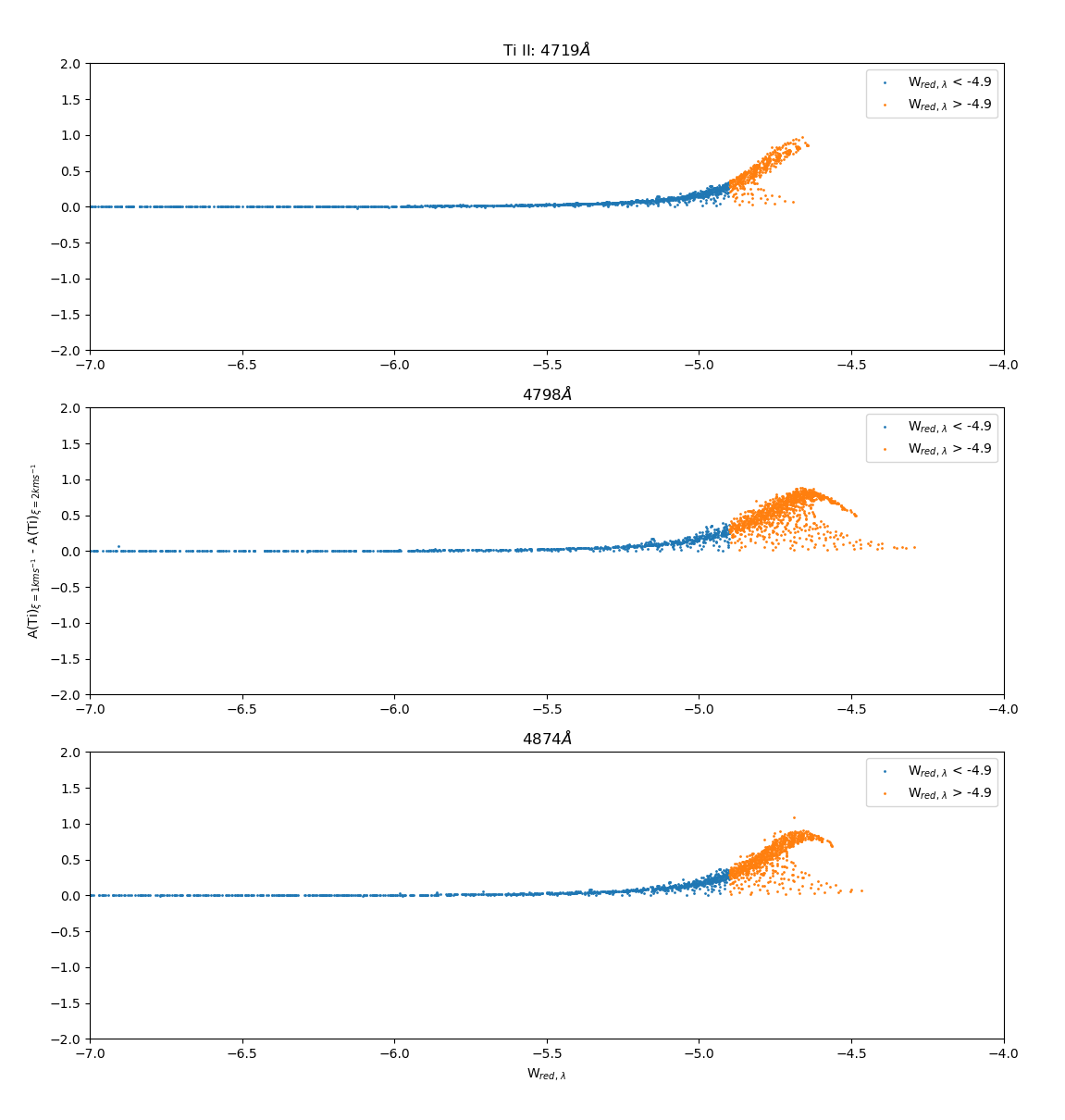}
  \caption{Difference in abundance for GALAH stars with microturbulences replaced by $\xi = 1$ and $\xi = 2$ kms$^{-1}$ for Ti II.} 
      \label{fig:vmictest2}
\end{figure*}

This appendix contains additional figures where Figure \ref{nlte_spacesTi2} represents the non-LTE effects for lines as in \fig{nlte_spaces} but for \ion{Ti}{II}, and the impact titanium abundance has on non-LTE effects in benchmark stars. Figures \ref{fig:vmictest}-\ref{fig:vmictest2} shows the impact reduced equivalent width has on microturbulence sensitivity in GALAH stars with their microturbulences replaced with 1 and 2kms$^{-1}$. Lines above $-4.9$ were removed due to the saturation effect and microturbulence sensitivity, shown in orange.

\end{appendix}
\end{document}